\documentstyle[12pt,thmsa,sw20lart]{article}

\input tcilatex
\begin{document}

\title{Anomalous temperature dependences of the susceptibility for the
one-3d-electron cation}
\author{Z.Ropka$^a$, Michalski$^{a,b}$ and R.J.Radwanski$^{a,b}$ \\
$^a$Center for Solid State Physics, \'{s}w. Filip 5, 31-150 Krak\'{o}w. \\
$^b$Inst. of Physics, Pedagogical University, 30-084 Krak\'{o}w,\\
POLAND.}
\maketitle

\begin{abstract}
We have studied properties of the one-3d-electron cation (Ti$^{3+}$, V$^{4+}$
ions) under the action of the low-symmetry crystal field in the presence of
the spin-orbit coupling. Very anomalous temperature dependences of the
magnetic susceptibility $\chi (T)$ have been obtained as resulting from the
off-cubic lattice distortion provided the intra-atomic spin-orbit coupling
is taken into account. It indicates that, despite of the weakness of the
spin-orbit coupling, the one 3d electron in a solid cannot be treated as a
free S=1/2 spin. The same holds for compounds containing the Ti$^{3+}$ and V$%
^{4+}$ ions.
\end{abstract}

\section{Introduction}

The one-3d-electron cation occurs in such compounds as CaV$_4$O$_9,$ MgVO$_3$%
, LiV$_2$O$_4$, LaTiO$_3$, NaV$_2$O$_5$, (VO$_2$)$_2$P$_2$O$_5$ that become
nowadays very popular [1-5]. Compounds containing V$^{4+}$ and Ti$^{3+}$
ions are usually treated as a S=1/2 system i.e. with the spin-only
magnetism. The neglect in the current literature of the orbital moment is
consistent with the widely-spread conviction that the orbital magnetism
plays rather negligible role due to the quenching of the orbital moment for
3d ions. However, this S=1/2 behavior is drastically violated in the above
mentioned compounds [1-5]. One of this drastic violation
experimentally-observed is associated with substantial departure of the
temperature dependence of the paramagnetic susceptibility from the Curie law
at low temperatures marked by the anomalous temperature dependence of the
magnetic susceptibility exhibiting a pronounced maximum centered at 20-100 K
and low susceptibility at low temperatures.

The aim of this paper is to show that anomalous temperature dependences of
the paramagnetic susceptibility can result from low-symmetry crystal-field
(CEF)\ interactions of the one-3d-electron cation provided the intra-atomic
spin-orbit (s-o) coupling is taken into account.

\section{Theoretical outline}

We have considered the electronic structure of the one-3d-electron cation
under the action of the crystal-field interactions H$_{CF}$ and the
spin-orbit coupling Hamiltonian H$_{s-o}$ as resulting from the
single-ion-like Hamiltonian [6-9]

\begin{center}
$H_d=H_{CF}+H_{s-o}=B_4(O_4^0+5O_4^4)+\lambda _{s-o}L\cdot S+B_2^0O_2^0+\mu
_B(L+$g$_eS)\cdot B_{ext}(1)$
\end{center}

The crystal field has been divided into the cubic part, usually dominant in
case of compounds containing 3d ions, and the off-cubic distortion written
by the second-order leading term B$_2^0$O$_2^0$. The last term allows the
influence of the external magnetic field to be calculated. $g_e$ amounts to
2.0023. It is necessary for calculations, for instance, of the paramagnetic
susceptibility - in fact the paramagnetic susceptibility is calculated [10]
as the magnetization in an external field of, say, 0.1 T.

The detailed form of the Hamiltonian (1) is written in the LS space that is
the 10 dimensional spin-orbit space. The L and S quantum numbers for one 3d
electron are equal to S=1/2 and L=2. The Hamiltonian (1) is customarilly
treated by perturbation methods [6] owing to the weakness of the s-o
coupling for the 3d ions in comparison to the strength of the crystal-field
interactions. We have accepted the weakness of the s-o coupling, what is
reflected by the sequence of terms in the Hamiltonian (1), but we have
performed direct calculations treating all shown terms in the Hamiltonian
(1) on the same foot. The separate figures, if presented are shown for the
illustration reasons.

Calculations [10] have been performed for physically relevant values of $%
\lambda _{s-o}$ close to +220 K (=150 cm$^{-1}$) - such the value is given
for the Ti$^{3+}$ ion [6 p.399]. The cubic CEF parameter B$_4$ is taken as
+200 K. The positive sign refers to the octahedral ligand surrounding of the
3d cation. For these parameters the overall effect of the s-o coupling
amounts to about 30 meV what is only 1.5 \% of the octahedral CEF splitting.
It turns out that despite of so small s-o coupling its effect on properties
is enormous, on the magnetic properties, in particular. It is related with
the fact of almost perfect compensation of the spin moment by the orbital
moment for the realized ground state.

\section{Results and discusion}

In Fig. 1 the temperature dependence of the paramagnetic susceptibility $%
\chi (T)$ is shown for different strength of the s-o coupling and different
off-cubic distortions. The inclusion of the s-o coupling breaks down the
Curie law at low temperatures as one can see comparing the curve 1 (no s-o
coupling) with curves 2 and 3, where $\lambda _{s-o}$ amounts to +220 K and
+100 K, respectively. In contrary to the curve 1, resembling the Curie law
expected for the free S=1/2 spin, the susceptibility shown by curves 2 and 3
exhibits at low temperatures a saturated behaviour resembling the Pauli
paramagnetism. The low-temperature saturation value decreases with
increasing the strength of the s-o copling. For $\lambda _{s-o}$ of +220 K
it saturates at level of 4$\cdot 10^{-3}$ $\mu _B/T$ per d-ion. It
corresponds to 2.2$\cdot 10^{-3}$ emu/mole provided that all spins in the
lattice contribute identically to the macroscopically observed
susceptibility. Values of this order are experimentally observed, indeed
[1-5]. It can be noted that despite of very anomalous behavior at low
temperatures, all $\chi (T)$ curves converge at temperatures above 300 K.
Above this temperature the temperature dependence can be approximated by the
Curie-Weiss law though the non-linearity of the $\chi ^{-1}(T)$ plot is
visible if too large temperature interval is considered.

The origin of such interesting temperature behavior of the paramagnetic
susceptibility is related with the low-energy electronic structure. In
Fig.2c the discrete low-energy electronic structure, computed for the
octahedral crystal-field parameter B$_4$ = +200 K and the spin-orbit
coupling constant $\lambda _{s-o}$ =+220 K is shown. Fig. 2d shows further
splitting into three Kramers doublet states (A,B,C) by the tetragonal
distortion of B$_2^0$=+10 K. It is obvious that the existence of such the
discrete energy spectrum will affect enormously electronic and magnetic
properties, in particular at low temperatures. For magnetic properties the
existence of the Kramers doublet ground state with the very small magnetic
moment and of the substantial moment at the excited state is extremally
important. With increasing temperature the excited states become thermally
populated and owing to drastically different values of the magnetic moment
the conventional Boltzmann distribution function produces the maximum in the 
$\chi (T)$ plot. In Fig. 3 the population of the three lowest states (A,B,C)
from Fig. 2d is shown.

We would like to point out that the exotic properties are not associated
with the shown parameters. Surely they do not result from a special choice
of parameters. Very similar low-energy structure and the $\chi (T)$ curve is
obtained for other parameters - of course the energy separations and the
magnetic moments will be different. We point out that the disccussed by us
parameters are physical measurable parameters and they are related with the
atomic physics and the local symmetry.

Very important outcome is that the 3d$^1$ electron system in the octahedral
oxygen surrounding, a MO$_6$ complex, has a very weakly-magnetic ground
state. This weakly-magnetic ground state results from almost perfect
compensation of the orbital and spin moments. We have calculated that the
ground state of the scheme shown in Fig. 2d has S=$\pm $0.50 and L=$\mp $%
1.00 and its Kramers-like eigenfunction is given as (the z component of L
and S are shown)

$\psi _o=0.9999\left| \pm 1,\mp \frac 12\right\rangle -0.01\left| 0,\pm
\frac 12\right\rangle $

where the sign $\pm $ refers to two conjugate Kramers states.

The eigenfunction of the first excited Kramers-like doublet is given as

$\psi _1=0.72\left| xy,\mp \right\rangle +0.68\left| \pm 1,\pm \frac
12\right\rangle ,$

where $\left| xy,\mp \right\rangle $ function has the usual meaning in the
CEF\ theory for the 3d cubic states [6-9] extended for the spin component as 
$\left| xy,\mp \right\rangle $ $=$ $\sqrt{1/2\text{ }}\left( \left| 2,\mp
\frac 12\right\rangle -\left| 2,\mp \frac 12\right\rangle \right) .$

Making use of the other functions orbital functions  $\left| xz\right\rangle 
$ and $\left| yz\right\rangle $ the eigenfunctions $\psi _o$ and $\psi _1$%
can be written as

$\psi _{o+}=\left| xz,-\right\rangle $

$\psi _{o-}=\left| yz,+\right\rangle $

$\psi _{1+}=$ $0.72\left| xy,-\right\rangle $ $+0.68\left| +1,+\frac
12\right\rangle $,

$\psi _{1-}=0.72\left| xy,+\right\rangle +0.68\left| -1,-\frac
12\right\rangle $.

From the shape of these functions one can easily see that for the functions $%
\psi _{o+}$ and $\psi _{o-}$ spin and orbital moments cancels each other.
For the $\psi _1$ functions the resultant moment equals $%
0.68^2(+3)-0.72^2(-1)$ i.e. 0.43 $\mu _B$ as is shown in Fig. 2d$.$

The formation of the very-weakly magnetic state the 3d$^1$ electron system
is really interesting result owing to the Kramers doublet ground state. The
compensation of the spin moment by the orbital moment obviously is related
with taking into account the intra-atomic spin-orbit coupling. The removal
of the Kramers-doublet ground-state degeneracy will produce further
interesting phenomena in ultra low temperatures. Their description goes,
however, beyond the present study.

\section{Conclusions.}

The very strong influence of the spin-orbit coupling and of the lattice
off-cubic distortions on the temperature dependence of the local
paramagnetic susceptibility has been revealed for the one-3d-electron cation
(the 3d$^1$ system). The highly-coupled spin-orbital 3d$^1$ system is
thought to be realized in Ti$^{3+}$ and V$^{4+}$ ions existing in ionic
compounds like CaV$_4$O$_9$, MgVO$_3$, LiV$_2$O$_4$, NaV$_2$O$_5$. These
off-cubic distortions cause very anomalous temperature dependences of the
local paramagnetic susceptibility resembling very much those experimentally
found in these presently-in-fashion compounds. We would like to point that
these anomalous $\chi (T)$ curves are obtained within the localized-electron
approach provided that the (weak) intra-atomic spin-orbit coupling is taken
into account. Although weak it has enormous influence on the low-energy
electronic structure and low-temperature magnetic and electronic properties.
We are aware that thecalculations have to be extended to take into account
other effects - first of them seems to be geometrical effects associated
with the non-collinearity of the local symmetry axes.The most important
result of this study is that the one 3d electron cannot be treated as the
free S=1/2 spin. The same holds for compounds containing the Ti$^{3+}$ and V$%
^{4+}$ ions.

{\bf Figure Captions }

Fig. 1. Calculated influence of the spin-orbit coupling and the lattice
off-cubic distortions on the temperature dependence of the paramagnetic
susceptibility $\chi $ (T) for the one-3d-electron cation (the 3d$^1$
system) in the dominant octahedral crystal field (B$_4$=+200K). 1) $\chi (T)$
in the presence of the octahedral CEF but in the absence of the spin-orbit
coupling - it shows the Curie law but with p$_{eff}$ of 2.29 $\mu _B$,
instead of 1.73 $\mu _B$ expected for the free S=1/2 spin; 2) $\chi (T)$ in
the presence of the octahedral CEF and the spin-orbit coupling $\lambda
_{s-o}$ = +220 K; 3) the same as 2) but for $\lambda _{s-o}$ of +100 K. 4,
5, 6) show the influence of the off-cubic lattice distortions in the
presence of the octahedral CEF (B$_4$=+200K) and the spin-orbit coupling $%
\lambda _{s-o}$ of +100 K; 4) with the tetragonal distortion B$_2^0$=+10 K;
5) and 6) with additional ortorhombic distortion B$_2^2$ of +5 K (5) and +10
K (6). In the inset the associated low-energy electronic structures are
schematically shown.

Fig. 2. The localized states of the one-3d-electron cation under the action
of the crystal field and spin-orbit interactions: a) the 10-fold degenerated 
$^2$D term realized in the absence of the CEF and the s-o interactions; b)
the splitting of the $^2$D term by the octahedral CEF\ surrounding B$_4$%
=+200 K ( $\lambda $ =0) yielding the $^2$T$_{2g}$ cubic subterm as the
ground state; c) the splitting of the lowest $^2$T$_{2g}$ cubic subterm by
the combined octahedral CEF and spin-orbit interactions ($\lambda $=+220 K, B%
$_4$=+200 K); the degeneracy and the associated magnetic moments are shown;
d) the further splitting by the tetragonal distortion of B$_2^0$=+10 K to
three Kramers doublet states (A,B,C); the degeneracy and the associated
magnetic moments are shown - the very small moment of the Kramers doublet
ground state and the substantial moment of the excited doublet is worth to
note.

Fig. 3. The temperature dependence of the population of the localized states
A, B and C shown in Figure 2d.


\begin{thebibliography}{99}
\bibitem{1}  M.P.Gelfand, Z.Weihong, R.R.P.Singh, J.Oitmaa and C.J.Hamer,
Phys.Rev.Lett. 77 (1996) 2794.

\bibitem{2}  S.Kondo et al. {\it Phys.Rev.Lett}, {\bf 78} (1997) 3729.

\bibitem{3}  S.K.Pati, R.R.P.Singh and D.I.Khomskii, Phys.Rev.Lett. 81
(1998) 5406.

\bibitem{4}  M.Isobe and Y.Ueda, J.Phys.Soc.Japan 65 (1996) 1178.

\bibitem{5}  S.Taniguchi, T.Nishikawa, Y.Yasui, Y.Kobayashi, M.Sato,
T.Nishioka, M.Kontani and K.Sano, J.Phys.Soc.Japan 64 (1995) 2758.

\bibitem{6}  A.Abragam and B.Bleaney, in: {\it Electron Paramagnetic
Resonance of Transition Ions}, (Clarendon Press, Oxford) 1970.

\bibitem{7}  C.Ballhausen, in: {\it Ligand Field theory }(McGraw-Hill, 1962)

\bibitem{8}  W.Low, in: {\it Paramagnetic Resonance in Solids} (Academic
Press, 1960).

\bibitem{9}  R.J.Radwanski,{\it \ Molecular Physics Reports} {\bf 15/16}
(1996) 113.

\bibitem{10}  the computer program is available on the written request to
the authors.
\end{thebibliography}
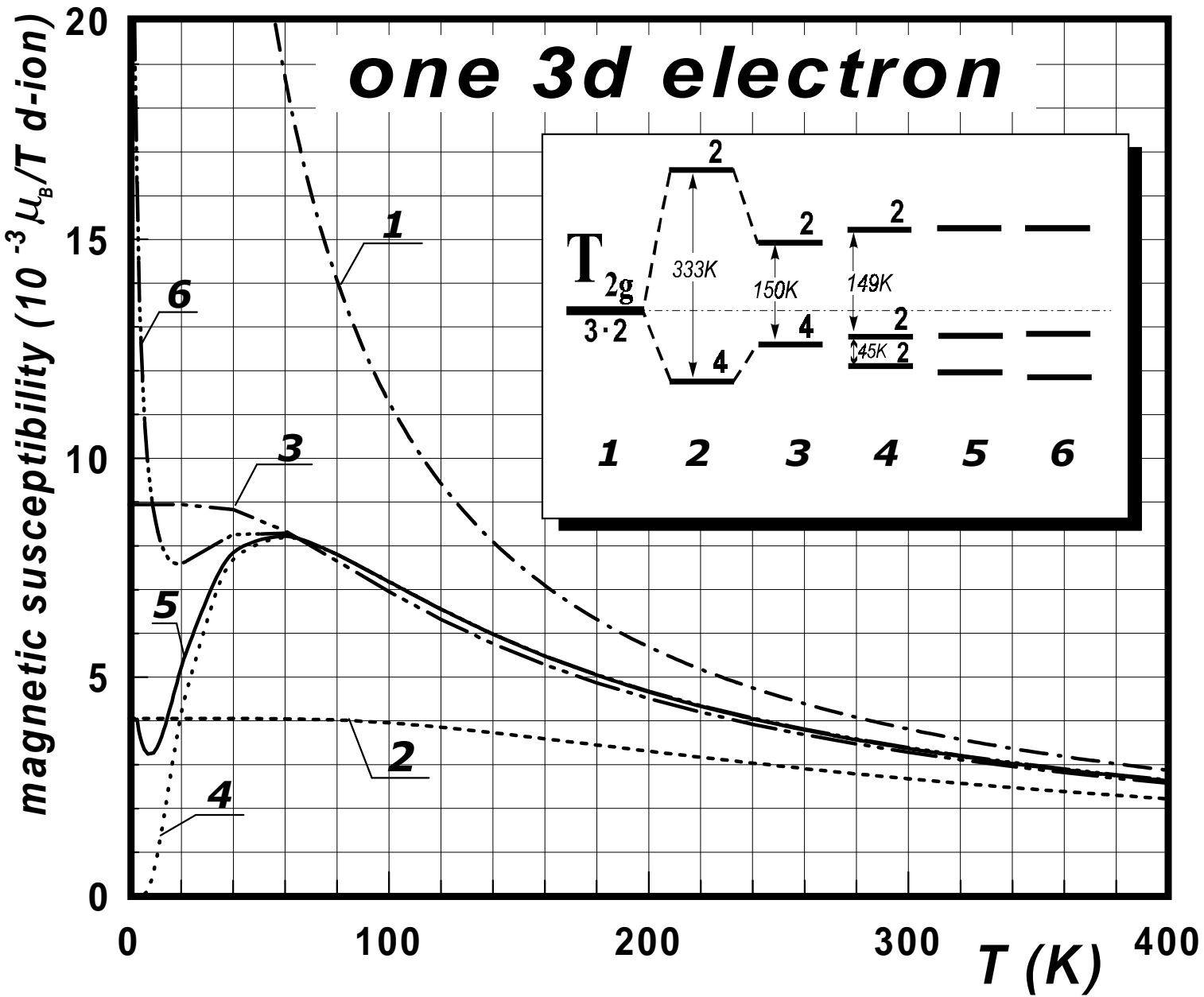
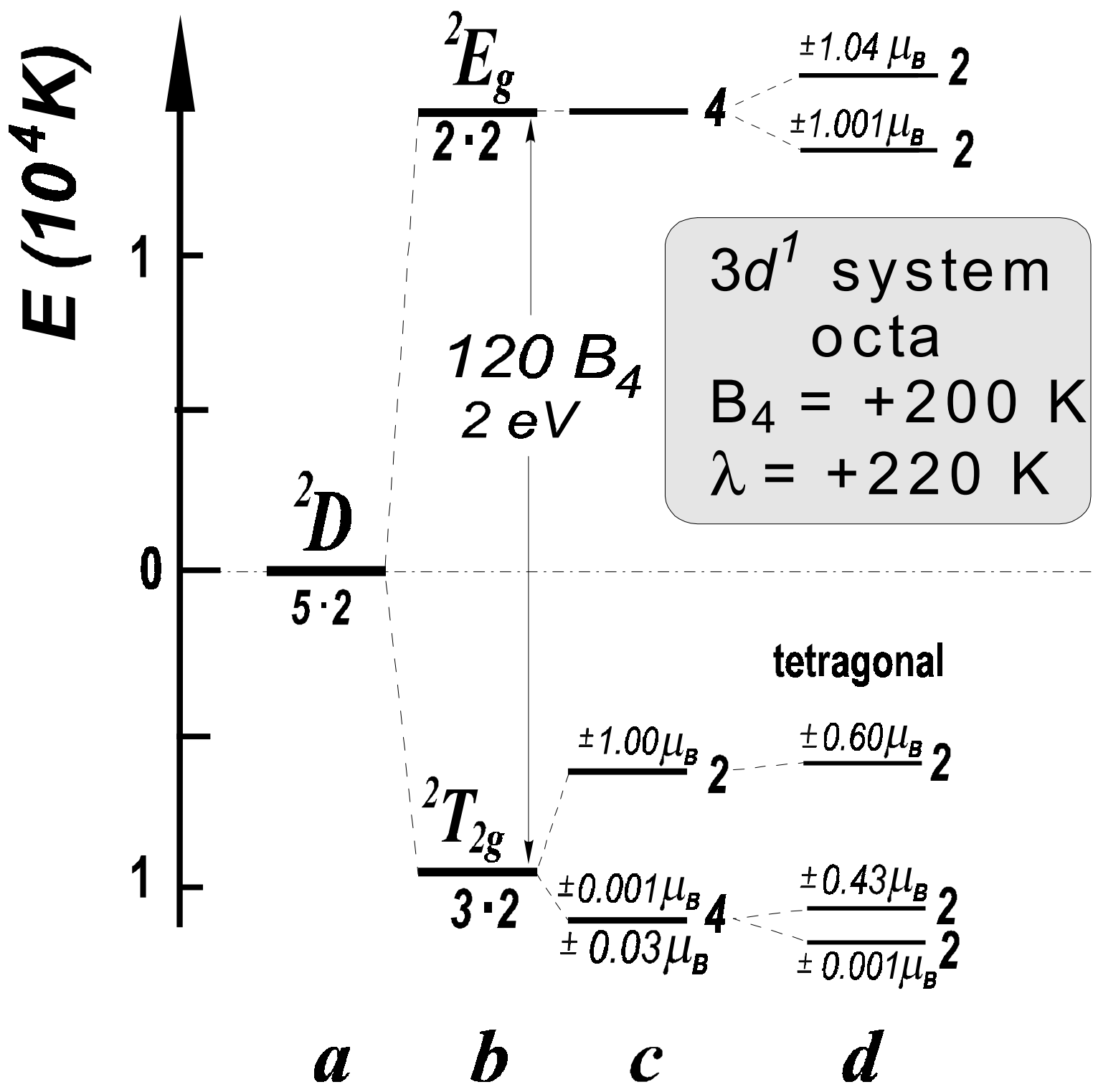
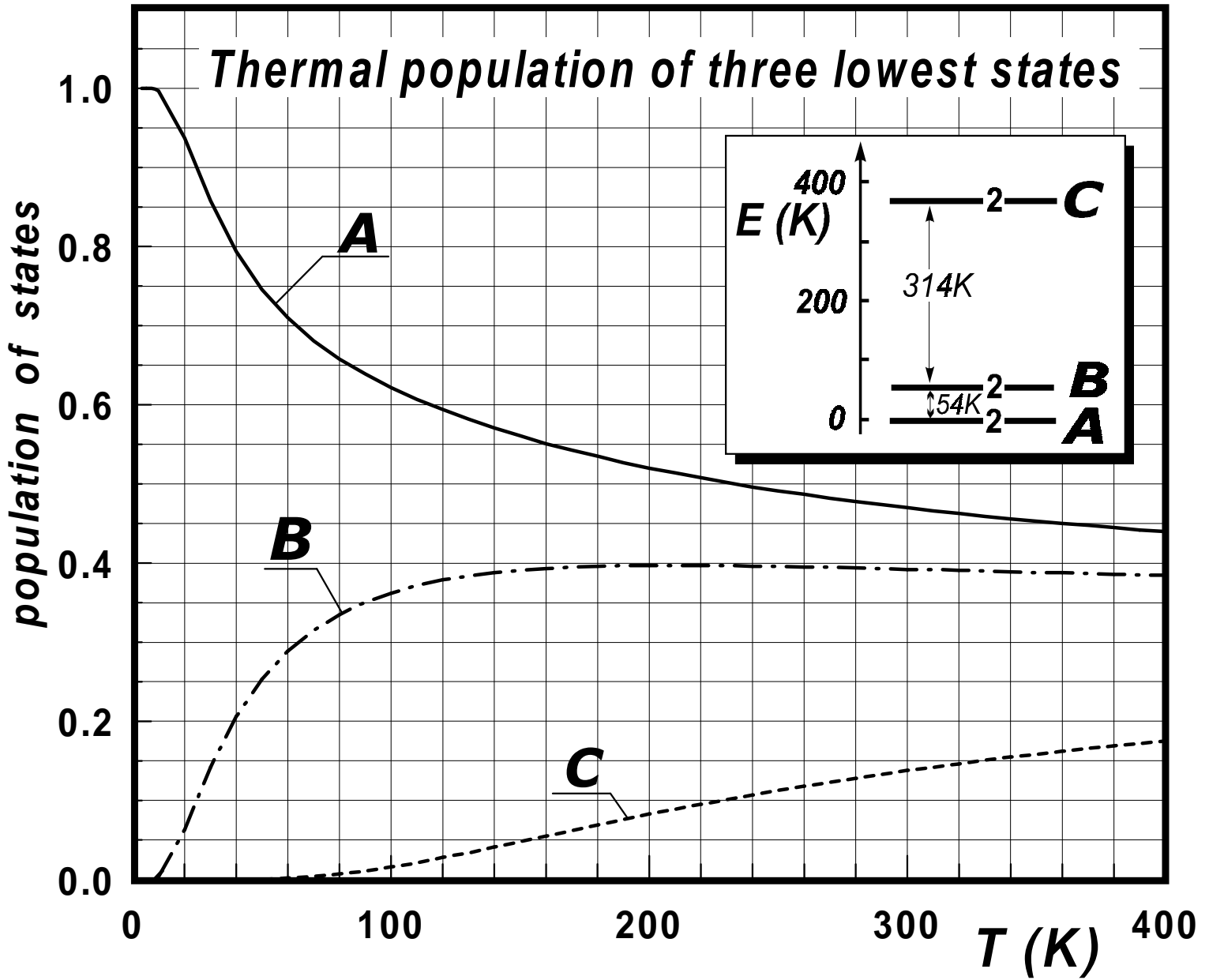
\end{document}